# Sequential Transmission Matrix Evaluation Via Spatiotemporal Transmitted Modes Decomposition

Shu Guo[1,2], Hao Zhang[2], Wenxue Li[1], and Lin Pang[1,2,*]

[1]*College of Physics, Sichuan University, Chengdu 610065, China*
[2] *LinOptx LLC, 6195 Cornerstone Court, San Diego, CA 92121, USA*
*\*pang@linoptx.com*

**Abstract:** The transmission matrix (TM) is a representation to describe the light scattering process through a scattering medium. The degree of control elements in TM is correlated with the capacity of evaluating enormous equations with tremendous number of unknow parameters, which seriously restricts the efficiency and accuracy of TM, and thus further confines applications such as image recovery. To completely remove this restriction, we propose decomposing TM and sequentially acquiring the dimension reduced decompositions regarding to the time and space invariance nature of transmitted field behind scattering medium. This proposed approach would not only have the ability to achieve high dimension transmission matrix with fully controllable elements, but also brings optimized signal-to-noise ratio during evaluation processing, which provides researchers a way to reach maximal focusing efficiency or image reconstruction resolution.

## 1. Introduction

Light passing through strong scattering medium, such as biological tissues, etc., experiences multiple scattering, which deteriorates the transmitted wave and thus generates a scramble of speckle patterns, making it difficult to transport predefined information such as optical images. Since the wavefront shaping approach was demonstrated to form focal spots behind the strong scattering medium [1,2], various methods were proposed to acquire the optimized wavefront, among them, directly presenting the scattering processing, transmission matrix (TM) attracts attention in wide applications, including image reconstruction [3,4], focusing light [5-9], exploit transmission eigenchannels [10-13], photonics length measurement [14].

However, current TM evaluation approaches require solving linear equations that correlate the predefined input fields and corresponding output fields. The ability to evaluate the input-output relations directly determines the dimension of the acquired TM, or say the number of modulation elements of TM, which is proportional to the focusing efficiency and resolution of the reconstructed image. As a result, in most reported applications, the dimension of TM is typically limited below 1024 modulation elements [15]. In order to acquire high dimension TM, great effort was made by Park's group, which processed the calculation of high dimension TM and data transition parallelly using two high-end PCs at the same time [16]. Another underlying issue in the current TM evaluation processing is oriented to information accuracy of the measured output responses, which strongly relies on the signal-to-noise ratio (SNR) during the evaluation processing. However fast the processing speed for TM evaluation approach is, low SNR is the main reason why the focusing enhancement via the TM evaluation approach is much lower than those achieved by other feedback based approaches [17-21].

In this paper, we propose an unconventional TM evaluation approach, which decomposes the evaluation of transmission matrix, to completely relax the strict hardware requirements for high dimension TM evaluation, while introducing the local optimization to increase the signal level when evaluating the output fields. The output field, regarding Huygens–Fresnel principle, is the linear superposition of individual transmitted fields, or incident modes, that are spatiotemporally decomposable. In other words, these transmitted fields are invariant in space and time in terms of their positions on the incident plane as well as the interference at the defined output site. The decomposition nature of spatiotemporal transmitted fields not only forms the foundation for the sequential evaluation of TM, but introduces the accumulation advantages to the evaluating decomposed regions on the output side during consecutive evaluations, which thus increases the SNR during the TM evaluation, and leading to superior efficiency of input field modulation, or more accurate representation of the scattering medium. The sequential decomposition TM evaluation approach is first described in detail based on space and time invariance of interference of the input modes, and then experimentally demonstrated.

## 2. Theory

According to the Huygens–Fresnel principle, the optical field behind the scattering medium on the output plane is the linear combination of all transmitted fields, or say, modulated input modes. When using a transmission matrix to describe the scattering processing, the output field is then determined by the input fields and the property of the scattering medium, which could be described as:

$$E_m^{\text{out}} = \sum_{i=1}^{N} t_{mi} E_i^{\text{in}} \tag{1}$$

where $E_i^{\text{in}}$ and $E_m^{\text{out}}$ are the complex amplitude fields on the input plane or the modulator plane, and on the output plane, respectively; the two items might also be called input modes and output modes. $N$ is the number of the input modes. The complex coefficients $t_{mi}$, represent the propagation property in the scattering system, which connects the $i$th input modes and $m$th output modes. The intensity of $m$th output mode is the magnitude of the field $E_m^{\text{out}}$:

$$\begin{aligned} I_m^{\text{out}} = \left|E_m^{\text{out}}\right|^2 &= \left(\sum_{i=1}^{N} t_{mi} E_i^{\text{in}}\right)^2 \\ &= \left(\sum_{i=1}^{N} \left(t_{mi} E_i^{\text{in}}\right)^2\right) + \sum_{i=1}^{N} \sum_{j=1, j\neq i}^{N} t_{mi} E_i^{\text{in}} \cdot t_{mj} E_j^{\text{in}} \end{aligned} \tag{2}$$

Here subscripts $i$ and $j$ indicate the indices of individual modes on the input plane. The [Eq. (2)] suggests that the intensity of output mode is the result of interference of all modulated input modes, which could be categorized into two groups. The first group is the superposition of the inner product of the modulated input modes themselves, and the second group represents a combination of cross product among individual modulated modes, which stands for the mutual interference of these modes. The [Eq. (2)] can be further regrouped regarding the locations on the input plane:

$$I_m^{\text{out}} = \left|E_m^{\text{out}}\right|^2 = \left(\sum_{i=1}^{N} t_{mi} E_i^{\text{in}}\right)^2$$

$$\left[\sum_{i=1, i\in Q_1}^{N_1} \left(t_{mi} E_i^{\text{in}}\right)^2 + \sum_{i=1, i\in Q_1}^{N_1} \sum_{j=1, j\neq i, j\in Q_1}^{N_1} t_{mi} E_i^{\text{in}} \cdot t_{mj} E_j^{\text{in}}\right.$$

$$\left. + \sum_{i=1, i\in Q_1}^{N_1} \sum_{j=1, j\in Q_2}^{N_2} t_{mi} E_i^{\text{in}} \cdot t_{mj} E_j^{\text{in}} + \ldots + \sum_{i=1, i\in Q_1}^{N_1} \sum_{j=1, j\in Q_k}^{N_z} t_{mi} E_i^{\text{in}} \cdot t_{mj} E_j^{\text{in}}\right] (3.1)$$

$$\left[\sum_{i=1, i\in Q_2}^{N_2} \left(t_{mi} E_i^{\text{in}}\right)^2 + \sum_{i=1, i\in Q_2}^{N_2} \sum_{j=1, j\neq i, j\in Q_2}^{N_2} t_{mi} E_i^{\text{in}} \cdot t_{mj} E_j^{\text{in}}\right. \quad (3)$$

$$\left. + \sum_{i=1, i\in Q_2}^{N_2} \sum_{j=1, j\in Q_1}^{N_1} t_{mi} E_i^{\text{in}} \cdot t_{mj} E_j^{\text{in}} + \ldots + \sum_{i=1, i\in Q_2}^{N_2} \sum_{j=1, j\in Q_k}^{N_z} t_{mi} E_i^{\text{in}} \cdot t_{mj} E_j^{\text{in}}\right] (3.2)$$

$$+ \ldots\ldots$$

$$\left[\sum_{i=1, i\in Q_k}^{N_z} \left(t_{mi} E_i^{\text{in}}\right)^2 + \sum_{i=1, i\in Q_k}^{N_z} \sum_{j=1, j\neq i, j\in Q_k}^{N_z} t_{mi} E_i^{\text{in}} \cdot t_{mj} E_j^{\text{in}}\right.$$

$$\left. + \sum_{i=1, i\in Q_k}^{N_z} \sum_{j=1, j\in Q_1}^{N_1} t_{mi} E_i^{\text{in}} \cdot t_{mj} E_j^{\text{in}} + \ldots + \sum_{i=1, i\in Q_k}^{N_z} \sum_{j=1, j\in Q_{k-1}}^{N_{z-1}} t_{mi} E_i^{\text{in}} \cdot t_{mj} E_j^{\text{in}}\right] (3.k)$$

The total $N$ input modes are now divided into segments $Q_1, Q_2, \ldots, Q_k$, with corresponding input modes numbers of $N_1, N_2, \ldots, N_z$. In [Eq. (3)], each square bracket stands for the interference contributions from one divided segment on the input plane, to the intensity on $m$th output mode. Take [Eq. (3.1)] as an example, the first term $\sum_{i=1, i\in Q_1}^{N_1} \left(t_{mi} E_i^{\text{in}}\right)^2$ corresponds to the intensity contribution of each mode inside the segment $Q_1$, then the second term $\sum_{i=1, i\in Q_1}^{N_1} \sum_{j=1, j\neq i, j\in Q_1}^{N_1} t_{mi} E_i^{\text{in}} \cdot t_{mj} E_j^{\text{in}}$ denotes the contribution of the interference between the input modes inside this segment, and till the third term $\sum_{i=1, i\in Q_1}^{N_1} \sum_{j=1, j\in Q_2}^{N_2} t_{mi} E_i^{\text{in}} \cdot t_{mj} E_j^{\text{in}} + \ldots + \sum_{i=1, i\in Q_1}^{N_1} \sum_{j=1, j\in Q_k}^{N_z} t_{mi} E_i^{\text{in}} \cdot t_{mj} E_j^{\text{in}}$ in the second column, it corresponds to the interference effect between the input modes inside $Q_1$ and those outside segment $Q_1$. In other words, [Eq. (3)] reminds us that the intensity of output mode could be evaluated sequentially by input modes.

For easy to follow, supposes that the modes on input plane are divided into four segments as shown at the most left column in Fig. 1, then [Eq. (3)] could be simplified as [Eq. (S1)] in Supplementary Document, the total $N$ modes in input plane are grouped to four regions $Q_1$, $Q_2$, $Q_3$, $Q_4$, each contains $N_1$, $N_2$, $N_3$, $N_4$ modes, respectively. In [Eq. (S1)], the four brackets represent the interference that are assigned to input modes from these four segments. Typically, in [Eq. (S1.1)], the first term indicates the intensity contribution from each mode inside region $Q_1$ with $N_1$ input modes. The second term then represents the interference contribution between those input modes inside the region $Q_1$. The contribution of the interference between the input modes inside the region $Q_1$ and input modes in region $Q_2$ is represented by the third term, so on and so forth, the fourth term and fifth term relate the interference contribution between $Q_1$ and region $Q_3$, and region $Q_4$, respectively. Similar to

[Eq. (S1.1)], the terms in [Eq. (S1.2)], [Eq. (S1.3)] and [Eq. (S1.4)] in [Eq. (S1)] then stand for the corresponding interference contributions in regions $Q_2$, $Q_3$ and $Q_4$, respectively.

At the desired position on the output plane, when the transmitted modes are modulated in such a way that interference is constructively achieved, or say all the optimized input modes turn to in phase conditions, the maximal concentration power at the output mode can be realized. In a nutshell, the point optimization is excited, this situation is described in [Eq. (4)], in which superscript of 'opt' indicates the optimized status.

$$
\begin{aligned}
I_m^{out} = \left| E_m^{out} \right|^2 &= \left( \sum_{i=1}^{N} t_{mi} E_i^{in,opt} \right)^2 \\
&= \Bigg[ \sum_{i=1, i\in Q_1}^{N_1} \left( t_{mi} E_i^{in,opt} \right)^2 + \sum_{i=1, i\in Q_1}^{N_1} \sum_{j=1, j\neq i, j\in Q_1}^{N_1} t_{mi} E_i^{in,opt} \cdot t_{mj} E_j^{in,opt} + \sum_{i=1, i\in Q_1}^{N_1} \sum_{j=1, j\in Q_2}^{N_2} t_{mi} E_i^{in,opt} \cdot t_{mj} E_j^{in,opt} \\
&\quad + \sum_{i=1, i\in Q_1}^{N_1} \sum_{j=1, j\in Q_3}^{N_3} t_{mi} E_i^{in,opt} \cdot t_{mj} E_j^{in,opt} + \sum_{i=1, i\in Q_1}^{N_1} \sum_{j=1, j\in Q_4}^{N_4} t_{mi} E_i^{in,opt} \cdot t_{mj} E_j^{in,opt} \Bigg] (4.1) \\
&\quad + \Bigg[ \sum_{i=1, i\in Q_2}^{N_2} \left( t_{mi} E_i^{in,opt} \right)^2 + \sum_{i=1, i\in Q_2}^{N_2} \sum_{j=1, j\neq i, j\in Q_2}^{N_2} t_{mi} E_i^{in,opt} \cdot t_{mj} E_j^{in,opt} + \sum_{i=1, i\in Q_2}^{N_2} \sum_{j=1, j\in Q_1}^{N_1} t_{mi} E_i^{in,opt} \cdot t_{mj} E_j^{in,opt} \\
&\quad + \sum_{i=1, i\in Q_2}^{N_2} \sum_{j=1, j\in Q_3}^{N_3} t_{mi} E_i^{in,opt} \cdot t_{mj} E_j^{in,opt} + \sum_{i=1, i\in Q_2}^{N_2} \sum_{j=1, j\in Q_4}^{N_4} t_{mi} E_i^{in,opt} \cdot t_{mj} E_j^{in,opt} \Bigg] (4.2) \\
&\quad + \Bigg[ \sum_{i=1, i\in Q_3}^{N_3} \left( t_{mi} E_i^{in,opt} \right)^2 + \sum_{i=1, i\in Q_3}^{N_3} \sum_{j=1, j\neq i, j\in Q_3}^{N_3} t_{mi} E_i^{in,opt} \cdot t_{mj} E_j^{in,opt} + \sum_{i=1, i\in Q_3}^{N_3} \sum_{j=1, j\in Q_1}^{N_1} t_{mi} E_i^{in,opt} \cdot t_{mj} E_j^{in,opt} \\
&\quad + \sum_{i=1, i\in Q_3}^{N_3} \sum_{j=1, j\in Q_2}^{N_2} t_{mi} E_i^{in,opt} \cdot t_{mj} E_j^{in,opt} + \sum_{i=1, i\in Q_3}^{N_3} \sum_{j=1, j\in Q_4}^{N_4} t_{mi} E_i^{in,opt} \cdot t_{mj} E_j^{in,opt} \Bigg] (4.3) \\
&\quad + \Bigg[ \sum_{i=1, i\in Q_4}^{N_4} \left( t_{mi} E_i^{in,opt} \right)^2 + \sum_{i=1, i\in Q_4}^{N_4} \sum_{j=1, j\neq i, j\in Q_4}^{N_4} t_{mi} E_i^{in,opt} \cdot t_{mj} E_j^{in,opt} + \sum_{i=1, i\in Q_4}^{N_4} \sum_{j=1, j\in Q_1}^{N_1} t_{mi} E_i^{in,opt} \cdot t_{mj} E_j^{in,opt} \\
&\quad + \sum_{i=1, i\in Q_4}^{N_4} \sum_{j=1, j\in Q_2}^{N_2} t_{mi} E_i^{in,opt} \cdot t_{mj} E_j^{in,opt} + \sum_{i=1, i\in Q_4}^{N_4} \sum_{j=1, j\in Q_3}^{N_3} t_{mi} E_i^{in,opt} \cdot t_{mj} E_j^{in,opt} \Bigg] (4.4)
\end{aligned} \quad (4)
$$

[Eq. (4)] shows that the light focusing at the desire locations on the output plane could be formed when all the input modes in region $Q_1$ are modulated to constructively interfere with themselves and those modes locate outside region $Q_1$, which is same held for input modes in region $Q_2$, $Q_3$ and $Q_4$. This condition implies that if the physical system is stable, optimal status in [Eq. (4)] could be achieved in sequence, bases on processing procedure both spatially and temporally. In other words, maximal intensity can be achieved in space sequential fashion, or one-by-one term in [Eq. (4)].

The in phase situation could be satisfied in such a way that, the first and second terms in the [Eq. (4.1)] are evaluated by conducting standard Hadamard matrix based TM evaluation process in region $Q_1$ to acquire its corresponding TM, named sub-TM and termed as $\text{TM}_{Q_1}^{\text{sub}}$, this process can be visualized at the row (a) in part1 of Fig. 1. We then come to [Eq. 4.2)], if input modes are turned off in region $Q_1$ during the TM evaluation process that corresponds to region $Q_2$, the resulted $\text{TM}_{Q_2}^{\text{sub}}$ only represents the scattering property related to region $Q_2$, and corresponds to first and second term in [Eq. 4.2)]. However, if TM evaluation process in region

$Q_2$ is conducted with optimized inputs in $Q_1$, represented by $\varphi_{Q_1}^{opt}$ derived from the acquired $\text{TM}_{Q_1}^{sub}$, are turned on, this would suggest the conduct of first three terms in Eq. (4.2), in which interference contribution from modes in $Q_1$ is considered. Consequently, the acquired sub-TM corresponds the region $Q_2$ is no longer just $\text{TM}_{Q_2}^{sub}$, but rather $\text{TM}_{Q_2 \text{ with } Q_1 \text{ on}}^{sub}$, which means that the first, second and third terms in [Eq. (4.2)] are taken into account, as visualized at row (b) in Fig. 1. Following the same procedures, TM evaluation process conducted in region $Q_3$ with optimization input from region $Q_1$ and $Q_2$ turned on would thus obtain $\text{TM}_{Q_3 \text{ with } Q_1, Q_2 \text{ on}}^{sub}$, which corresponds to first four terms in [Eq. (4.3)]. Similarly, $\text{TM}_{Q_4 \text{ with } Q_1, Q_2, Q_3 \text{ on}}^{sub}$ in region $Q_4$ with turned on optimized inputs from region $Q_1, Q_2$ and $Q_3$ would be acquired, which corresponds to the realization of all the terms in [Eq. (4.4)]. After conducting the TM evaluation process from $Q_1$ to $Q_4$ one by one while keeping optimization from measured regions turned on, the so called sub-TMs are determined:

$$\text{TM}_{Q_1}^{sub}, \text{TM}_{Q_2 \text{ with } Q_1 \text{ on}}^{sub}, \text{TM}_{Q_3 \text{ with } Q_1, Q_2 \text{ on}}^{sub}, \text{TM}_{Q_4 \text{ with } Q_1, Q_2, Q_3 \text{ on}}^{sub}$$

However, the assemble of regional sub-TMs on the four regions do not represent the actual TM for entire input modes, because the last three terms in [Eq. (4.1)], the last two terms in [Eq. (4.2)], and the last term in [Eq. (4.3)], have not been evaluated yet. As a result, the sub-TMs correspond to region $Q_1, Q_2, Q_3$ should be measured again with all the measured sub-TMs are applied to generate optimal fields. After completing another iteration that is shown row 5 to row 8 in Fig. S1 in the Supplementary Document, the sub-TMs such as $\text{TM}_{Q_1 \text{ with } Q_2, Q_3, Q_4 \text{ on}}^{sub}$, $\text{TM}_{Q_2 \text{ with } Q_1, Q_3, Q_4 \text{ on}}^{sub}$, $\text{TM}_{Q_3 \text{ with } Q_1, Q_2, Q_4 \text{ on}}^{sub}$, $\text{TM}_{Q_4 \text{ with } Q_1, Q_2, Q_3 \text{ on}}^{sub}$ would be gradually obtained. Consequently, all terms in four brackets in [Eq. (4)] could be realized, and the ultimate TM, the presentation of the scattering medium, corresponding to the entire input region is accessed.

As described above, sequentially processing the sub-TM evaluation procedures one by one, while turning on the optimized input modes from the previous evaluated regions, the TM with large number modulation elements could be realized. This proposed approach offers a chance to build maximally sizable TM based on all available modulation pixels on the incident plane, or modulator, without occupying excess computation power. It also provides an accessible way to acquire the TM of the scattering medium with highest efficiency in many applications like enabling the maximal energy delivery and high quality imaging through the scattering medium.

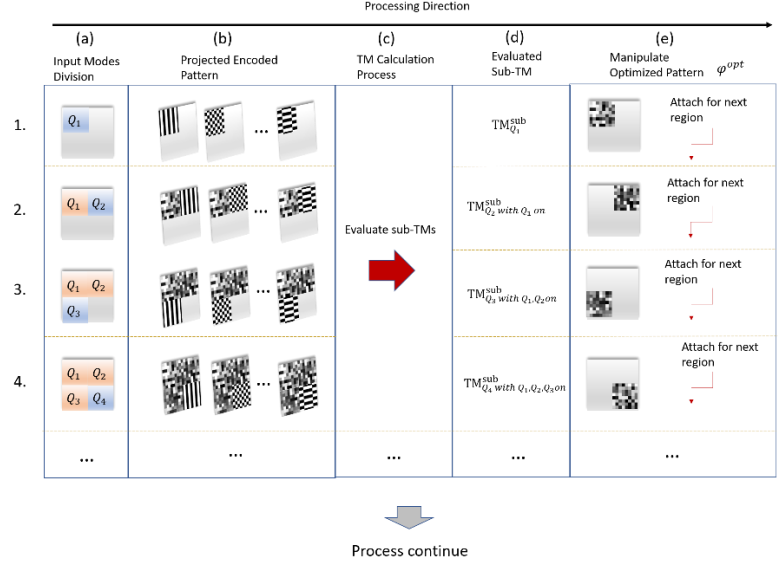

Fig. 1. Process of sequential TM measurement. The input modes are regrouped to four regions as $Q_1$, $Q_2$, $Q_3$, $Q_4$ in (a) frame. In row 1, the input modes in region $Q_1$ are exploiting to display encoded Hadamard basis that same as the size of $Q_1$ in (b) frame, the corresponding sub-TM is then calculated according to received intensity signal in (c) and (d) frame, and the optimized distribution is extracted from measured sub-TM in (e) frame. In row 2, the optimized distribution in $Q_1$ is kept displaying while region $Q_2$ is projecting encoded Hadamard basis to scattering medium. As a result, the sub-TM $\text{TM}^{\text{sub}}_{Q_2 \text{ with } Q_1 \text{ on}}$ calculated in $Q_2$ includes interference from $Q_1$. In row 3, the procedure is same as row 1 and row 2, the derived optimized distribution from $Q_1$ and $Q_2$ are displaying on DMD as $Q_3$ manipulates Hadamard basis. The total evaluation process would lead to the roverary of entire TM when interference among all transmitted modes are considered after the later measurement from row 4 to row 8 in Fig. S1 in Supplementary Document.

## 3. Experiment Result

### 3.1 Demonstration of Sequential TM evaluation measurement: compare direct TM and sequential TM evaluation approaches

Experiments are conducted to acquire the transmission matrix via the proposed sequential TM evaluation that was described in section 2. For simplification, the modulator elements that indicate input modes is selected as $128 \times 128$, which could be divided into four segments with dimension of $64 \times 64$, each segment corresponds to $\text{TM}^{\text{sub},64 \times 64}$. The sequential TM evaluation is then compared with the direct TM evaluation approach, which measures $\text{TM}^{\text{dir},128 \times 128}$ with Hadamard matrix in size of $16384 \times 16384$. The digital micromirror device (DMD), which offers a high framerate, is introduced as the light modulator to conduct binary amplitude modulation [22]. To measure the TM of the scattering medium, we follow the amplitude modulation method that projects the encoded Hadamard basis [16,23]. The input modulation pattern on DMD is encoded as $\mathbf{v}_n^{\pm} = 1/2(\mathbf{h}_1 \pm \mathbf{h}_n)$, where $N \times 1$ vector $\mathbf{h}_n$ is $n$th column of Hadamard matrix $\mathbf{H}$ with dimension $N \times N$, while the first column $\mathbf{h}_1$ is adding as interference field. The corresponding intensity through the scattering medium is:

$$I_n^{\pm} = \left|\mathbf{T}\nu_n^{\pm}\right|^2 = \left|\frac{1}{2}\mathbf{T}(\mathbf{h}_1 \pm \mathbf{h}_n)\right|^2 = \frac{1}{4}|\sigma_1|^2 + \frac{1}{4}|\sigma_n|^2 \pm \frac{1}{2}\mathrm{Re}(\sigma_1^*\sigma_n) \tag{5}$$

Here $\mathbf{T} = [t_1 \ t_2 \ ... \ t_n]$ is the transmission matrix. $\sigma_1, \sigma_n$ represent the output field that corresponds to reference input $\mathbf{h}_1$ and $n$th column $\mathbf{h}_n$, the real part of $\sigma_n$ can be calculated by the corresponded three intensity signals:

$$\mathrm{Re}(\sigma_n) = \frac{I_n^+ - I_n^-}{\sqrt{I_1}} \tag{6}$$

$I_1$ indicates the intensity that corresponds to the input pattern $\mathbf{h}_1$. The measurement number of signal intensity to determine TM is defined as $3 \times N$. As all the corresponding output fields are determined, the transmission matrix which connects the input field and output field can be measured by:

$$\mathrm{Re}(\mathbf{T}) = \mathbf{SH} = \frac{1}{N}[\mathrm{Re}(\sigma_1) \ \mathrm{Re}(\sigma_2) \ ... \ \mathrm{Re}(\sigma_n)]_{1 \times N} \times [h_1 \ h_2 \ ... \ h_n]_{N \times N} \tag{7}$$

Where $\mathbf{S} = 1/N \cdot \mathrm{Re}(\sigma_1 \ \sigma_2 \ ... \ \sigma_N)$ corresponds to real value output field vector. TM could then be evaluated by solving [Eq. (7)], in which Hadamard matrix $\mathbf{H}$ is known, while $\mathbf{S}$ is just acquired. The optimized amplitude distribution which can achieve maximal transmitted intensity at the desired output position is generated according to the transmission coefficients in TM following the criteria:

$$e_n^{\mathrm{opt}} = \begin{cases} 1 & \mathrm{Re}(t_n) \geq 0 \\ 0 & \mathrm{Re}(t_n) < 0 \end{cases} \tag{8}$$

The symbol $e_n^{\mathrm{opt}}$ represents the elements of the input field, which is binary and able to generate a focal spot behind the scattering medium. Above TM evaluation algorithm is conducted to acquire the TM for both $\mathrm{TM}^{\mathrm{dir},128 \times 128}$ in traditional TM evaluation approach and $\mathrm{TM}^{\mathrm{sub},64 \times 64}$ in sequential TM measurement method.

The experiment setup is illustrated in Fig.2, in which a laser (Coherent Genesis MX) beam of 532nm is expanded by a $5\times$ microscopy and then collimated by L1. The angle of DMD surface is adjusted to satisfy blazed grating for 532nm wavelength illumination. The DMD surface is then imaged onto the front surface of the scattering medium (Thorlabs Ground Glass Diffuser 120 Grit) by a 4-f image system of lenses L2 and L3, A CCD camera (Thorlabs CS2100M) located behind the scattering medium, collects the scattered light at the numerical aperture $\mathrm{NA} = 0.0215$.

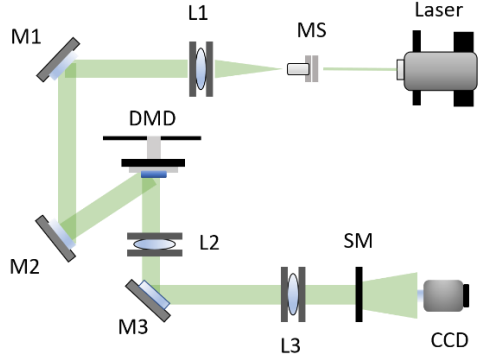

Fig 2. Experiment setup. Laser: Coherent Genesis MX 532nm; L1: lens, $f = 100mm$; L2: lens, $f = 200mm$; L3: lens, $f = 150mm$; CCD: Thorlab-2100M; MS: microscope (Newport M- 5×); M1~M5: mirror; SM: Thorlabs Ground Glass Diffuser 120 Grit. DMD: DLP700 ($1024 \times 768$).

First, $TM^{dir,128\times128}$ is measured with the encoded Hadamard matrix basis of $16384\times16384$, with total measurements of $16384\times3=49152$ conducted to calculate signal output field $\mathbf{S}$. $TM^{dir,128\times128}$ is then recovered by exploiting vector-matrix multiplication with dimension of $\mathbf{S}_{1\times16384}\mathbf{H}_{16384\times16384}$. When optimization distribution $\varphi^{opt,128\times128}$ shown in Fig3.(c) is derived from $TM^{dir,128\times128}$ and displayed on the DMD, the focal spot is formed at the region of interest as shown in Fig. 3(b). If all modulated pixels are set to 1 (pixel is set at 'on' state) as a plane wave input illumination, the intensity distribution on the CCD is a randomized speckle pattern, as shown in Fig. 3(a). The enhancement factor (EF), defined as the ratio of intensity at a focal spot to mean background intensity illuminated with plane wave (all pixels $128\times128=16384$ on DMD set as 'on' in our case), is measured as around 291.

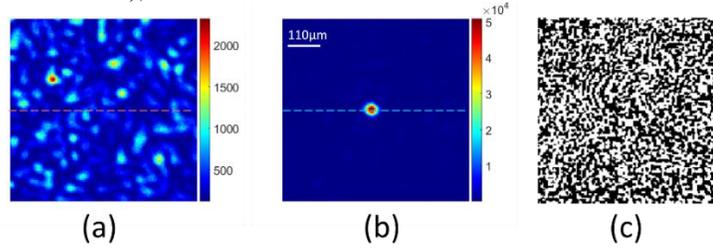

Fig. 3. Experiment result for direct TM evaluation. (a) Speckle distribution in the region of interest when $128\times128$ pixels on DMD turn to 1 (on) state. (b) Focal spot generated by optimized amplitude distribution; the blue dash line represents the cross-section along the maximum intensity point. (c) Optimized amplitude distribution $\varphi^{dir,128\times128}$ that is derived from $TM^{dir,128\times128}$, the value range is binary for DMD usage.

Then the proposed sequential evaluation process described in section 2 is conducted to evaluate $TM^{seq,128\times128}$, but by calculating sub-TMs termed $TM^{sub,64\times64}$ on the 4 regions $Q_1$, $Q_2$, $Q_3$, $Q_4$ as illustrated in Fig. 4. The sub-TM on region $Q_1$ is first evaluated as $TM^{sub,64\times64}_{Q_1}$ with encoded Hadamard basis in size of $4096\times4096$ by calculating $\mathbf{S}_{1\times4096}\mathbf{H}_{4096\times4096}$, which differs from the direct evaluation that requires $\mathbf{S}_{1\times16384}\mathbf{H}_{16384\times16384}$ computing volume. Fig.4(a1) shows the evaluated binary optimized amplitude distribution $\varphi_{Q_1}^{opt,64\times64}$ that derives from $\text{TM}_{Q_1}^{sub,64\times64}$, while Fig.4(a2) showing the corresponding intensity distribution as $\varphi_{Q_1}^{opt,64\times64}$ is displayed on the DMD. As suggested in section 2, following the same TM evaluation process on region $Q_2$ while displaying the $\varphi_{Q_1}^{opt,64\times64}$ on the $Q_1$ region of the DMD, the $\text{TM}_{Q_2 \text{ with } Q_1 \text{ on}}^{sub,64\times64}$ could be evaluated, and the corresponding optimize distribution $\varphi_{Q_2 \text{ with } Q_1 \text{ on}}^{opt,64\times64}$ is shown in Fig.4(b1). The intensity distribution on output plane when $\varphi_{Q_1}^{opt,64\times64}$ and $\varphi_{Q_2 \text{ with } Q_1 \text{ on}}^{opt,64\times64}$ are displaying on region $Q_1$ and $Q_2$ of DMD is shown in Fig.4(b2). In accordance with the procedure in section 2, the sub-TMs of four regions are evaluated one by one, and the intensity distribution are respectively shown in Fig.4(a2) to Fig.4(g2) as all the optimized amplitude distributions derived from the previous sub-TMs are displayed on the corresponding regions of DMD, that are shown in Fig.4(a1) to Fig.4(g1). The maximum intensity value $I_{max}$ at the desired output position during each evaluation is listed in the Table S1 in Supplementary Document, with the sequentially acquired sub-TMs on the corresponding regions. For better visualization, $I_{max}$ is plotted in Fig.4(i) as well. The blue diamond shows experimental data value corresponds to Table S1. The monotonically increased red fitting curve indicates the accumulated effect of displayed optimized distribution. After using the determined $\varphi^{opt,64\times64}$ to combine the entire optimized distribution $\varphi^{opt,128\times128}$ that shown in Fig.4(h1) and display it to DMD, a bright focal spot is formed shown in Fig.4(h2), The corresponding $I_{max}$ is 65535, which is plotted as a black circle in Fig.4(i). The above sequential TM evaluation process was iterated another two times, and the resulting $I_{max}$ that led by their corresponding entire optimized distribution are 65535, 55672 respectively, as a red circle and green circle indicate, which means that the extended evaluation process does not improve the efficiency anymore. This is in accordance with the proposed principle that the sequential TM evaluation just requires two rounds of measurements. The residual discrepancy in the achieved intensities shown in Fig.4(i) might result from the perturbation of system as time consuming.

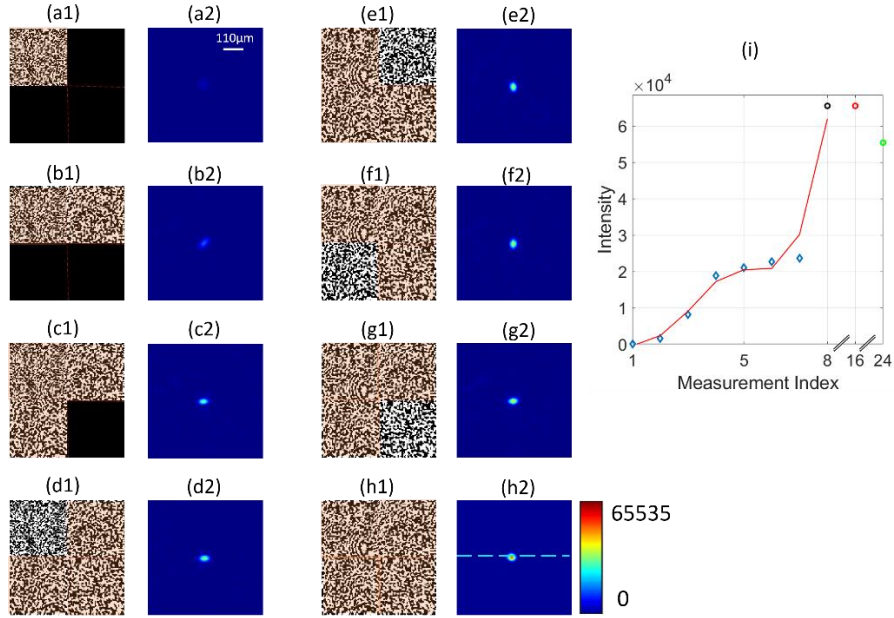

Fig. 4. Experiment result for sequential TM approach with size $128\times128$. (a1)-(g1) indicate optimized amplitude distribution during each sequential evaluation corresponds to divided region on DMD. Each pattern is in size of $128\times128$, the red line indicates the border of four divided $64\times64$ regions. (a2)-(g2) indicate intensity distribution in the region of interest on the output plane when the optimized distribution noted by orange covers in (a1)-(g1) are displaying on DMD, respectively. (h1) represents the eventual optimized distribution, the resulting intensity pattern is shown in (h2), the blue dash line indicates the cross-section along with $I_{max}$ position; (i) $I_{max}$ at desired output position during sequential TM measurement, the blue circle data points are corresponding to $I_{max}$ in pattern (h2)-(h2), the red curve indicates the fitting data of blue circles, the black circle represents the $I_{max}$ when eventual optimized distribution are displaying on DMD as shown in (h1) and (h2). Then the red and green circles indicate the results that repeat sequential TM evaluation for two times, respectively.

Comparison between direct TM evaluation and proposed sequential TM evaluation is given in Fig.5, in which the intensity distribution across the formed focal spot after applying the measured optimized amplitude distributions from both methods shown in Fig. 3(b) and Fig.4(h2). The corresponding mean background intensities that are illuminated with unmodulated plane wave for the two cases are also drawn in Fig. 5 as green line, from which the EF for $TM^{dir,128\times128}$ and $TM^{seq,128\times128}$ can be calculated as 291 and 362, respectively. The sequential TM measurement improves enhancement by 24.4% compared to direct TM measurement. The increase of efficiency can be attributed to improved signal to noise ratio (SNR) during the establishment of the sub-TMs, which will further be discussed later.

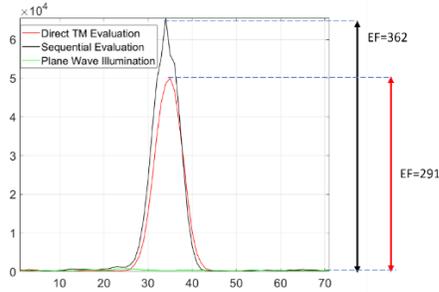

Fig. 5. Comparison of direct TM evaluation approach and sequential TM evaluation approach. The red curve and black curve represent cross section of the focal spot that is generated from direct TM evaluation and sequential TM evaluation. The green curve shows the corresponding cross section of speckle pattern that formed by plane wave illumination on DMD, the enhancement factors for two methods are visualized by black arrow line and red arrow line, each indicates the enhancement factor (EF) equal to 362 and 291.

*3.2 High Dimension Transmission Matrix acquisition*

Besides the improved efficiency or increased EF, the proposed sequential TM evaluation is also of great advantage over the traditional TM evaluation approach, for the reason that it avoids to solve large scale linear equations with tremendous unknown parameters for high dimension TM evaluation. Typically, the huge computation capacity requirement of traditional TM evaluation might be impossible for regular computing equipment. From the above demonstration, the sequential TM evaluation approach offers a method to acquire TM by sequentially evaluating small dimension $TM^{sub}$. Therefore, in principle, any high dimension TM could be readily evaluated at any operating system.

As an example, $TM^{seq,512\times512}$ with a dimension of $512\times512$ is acquired using sequential TM evaluation approach, which, however, would be extremely difficult to be conducted with regular computing power [16]. For the direct TM evaluation approach, the Hadamard matrix dimension will extend to $262144 \times 262144$, and the vector-matrix multiplication size of [Eq. (7)] then becomes $\mathbf{S}_{1 \times 262144} \mathbf{H}_{262144 \times 262144}$. But in contrast, in sequential approach, $\mathrm{TM}^{seq,512 \times 512}$ was also acquired with the sequential TM evaluation approach by evaluating sub-TMs with a size of $64 \times 64$ on sixty-four divided regions $Q_1, Q_2, ..., Q_{64}$. Fig.6(a) shows $I_{max}$ when the optimized amplitude distributions derived from previously acquired sub-TM on DMD. The intensity increases from 219, which corresponds to $\mathrm{TM}^{sub}_{Q_1}$, to 43795 after the 1st round evaluation. During the 2nd round evaluation, i.e., 65th measurement to 128th measurement in Fig. 6(a), still, the intensity increases from 43795. It can be saw that the rising rate of intensity value from the 1st measurement to roughly 30th measurement is flatter than 31st to around 64th measurement. The relatively lower raising rate from 1st to 30th measurement might result from the weak signal level, but as the ratio of modulation elements increases, the detected signal level from 31st to 64th measurement gradually improve from the noise level, and thus gains higher point optimization efficiency. $I_{max}$ is fluctuating from 65th to 128th measurement, which may be caused by mechanical vibration of the system. Fig.6(b) shows the optimized amplitude distribution $\varphi^{opt,512 \times 512}$ that is originated from $\mathrm{TM}^{seq,512 \times 512}$. The generated focal spot after the scattering medium is shown in Fig.6(c), and the corresponding logarithm intensity distribution is shown in Fig.6(d).

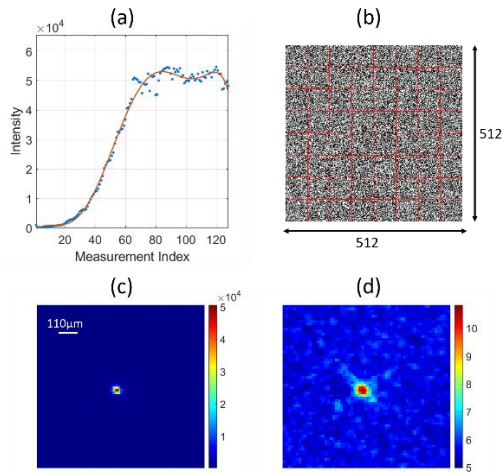

Fig.6 Experiment result of sequential TM measurement in size of $512 \times 512$. (a) $I_{max}$ at the desired position on output plane during sequential TM measurement in 64 regions that formed by optimized distribution from $\mathrm{TM}^{seq,512 \times 512}$, blue dots indicate original intensity value, while red curve fitting the trend of them; (b) Eventual optimized distribution $\varphi^{opt,512 \times 512}$ from sequential TM approach, the red line shows the division of $8 \times 8$ regions; (c) Focal spot at desired position when $\varphi^{opt,512 \times 512}$ is projected to the scattering medium; (d) Logarithm distribution of (c).

With the proposed sequential TM evaluation approach, different size of TM were readily acquired. The relationship between the EF of the focal spot and the number of the modulation elements of acquired TM is shown in Fig.7, in which five data points correspond to TM with size of $32\times32$, $64\times64$, $128\times128$, $256\times256$ and $512\times512$, respectively. The evaluation result of TM in size of $256\times256$ is shown in Fig. S2 in Supplementary Document. The corresponding EF value are $33\pm5$, $74\pm35$, $270\pm36$, $327\pm22$, $648\pm60$ averaged over three time experiments. The EF versus input modulation elements is fitted as:

$$\eta = \left\{1+\left(\frac{N}{2\alpha}-1\right)/\pi\right\}+\beta \tag{9}$$

$N$ represents the total number of control elements, $\alpha$ and $\beta$ are variable factors that depend on experiment environment. In our case, $\alpha=50$ and $\beta=40$. When the condition is ideal, the factor $\alpha$, $\beta$ should be neglected as EF only depends on the control elements number, for instance, as conducted in reference [22]. But in our study, experiments were processed in a certain level SNR environment, which is far insufficient to an ideal noise free system.

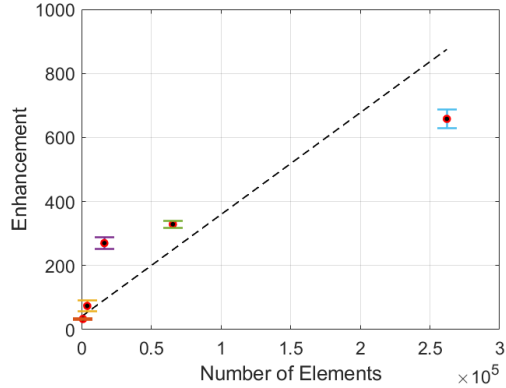

Fig. 7. The enhancement factor of focal spots versus modulation elements number at $32\times32$, $64\times64$, $128\times128$, $256\times256$ and $512\times512$. The error bar indicates deviation for 3 times experiment results, the black dash line represents fitting curve from [Eq. (9)].

## 4. Discussion

The acquired TM with size $512\times512$ covers $70\%$ of total modulation elements of the DMD (XGA) with resolution of $1024\times768$ pixels. Actually, the entire number of the pixels of the DMD could be modulated when the whole DMD modulation region is divided into $16\times12=192$ sub-regions with $64\times64$ elements for each, on which conducting sub-TM evaluation sequentially would give us the full control of each modulation elements on the DMD or other types spatial light modulator, and maximal enhancement could be achieved theoretically.

In the sequential TM evaluation approach, the selection of the size of the sub-TM is critical. In principle, there would be no limitation for the definition of the size of the sub-TM. For example, the XGA version with $1024\times768$ pixels could be divided into $16\times12=192$ regions for evaluation of $TM^{sub,64\times64}$ in size of $64\times64$, or $32\times24=768$ regions correspond to $TM^{sub,32\times32}$. Especially for those experiments that are confronted with the budget limit for the computing power, the reduced size of Hadamard matrix would bring fewer number of linear equations to be solved. For other formats DMD like WXGA that is equipped with a resolution of $1200\times800$, sequential TM measurement method has the ability to achieve the full modulation pixel control by dividing all elements into $75\times 50=3750$ regions with $\text{TM}^{\text{sub},16\times16}$. However, there are two factors are influential. First, the reduced size of sub-TM leads to increasing data loading, transition and experiment cycles, which potentially extends the total measurement time. Second, it would also affect the signal to noise ratio (SNR), measured as the output response versus the background noise variance. When the size of sub-TM is reduced, the decreasing number of elements results in the attenuation of illumination power through scattering medium. In matter of this situation, if the corresponding intensity at the output plane reduces to the background noise level, more error would be introduced into the evaluated TM, and thus leading to lower EF. In fact, we also acquired $\text{TM}^{\text{seq},128\times128}$ by sequential TM evaluation with sub-TM of $32\times32$. The eventual enhancement of focal spot is roughly same as the experiment from TM acquirement with sub-TM of size $64\times64$, which means that SNR level is sufficient for conducting sub-TM with size of $32\times32$ in our current experimental conditions. However, as we further reduced the size of sub-TM to $8\times8$, the overall SNR decreased sharply. On the other hand, the intensity at the desired output position might be saturated under a certain detector dynamic range when previously evaluated amplitude distributions are displayed on the corresponding regions. To overcome this limitation, one could decrease the exposure time or reduce the incident power; but still, the overall SNR would be sacrificed.

One of the advantages of the proposed sequential TM evaluation approach is that it enables more accurate representation of the scattering medium, which gives rise to higher efficiency regarding to the focusing after displaying the derived optimized amplitude distribution on the DMD, comparing with the traditionally acquired TM with the same control modulation elements. As we demonstrated in section 3.1, the sequentially acquired $\text{TM}^{\text{seq},128\times128}$ presents EF of 362 over that of 291 for $\text{TM}^{\text{dir},128\times128}$, showing 24.4% improvement. The reason for the better presentation accuracy of TM is that we overcome the insufficient SNR issue that appears in the direct TM evaluation approach when establishing the input-output relationship in the experiments. Accurately acquiring the corresponding output fields is the key to evaluate the TM that precisely represents the scattering medium, which is affected by SNR that is defined by the measured intensity at the desired target position ($I_d$) over the average background intensity $\langle I_0 \rangle$:

$$\text{SNR} = \frac{I_d}{\langle I_0 \rangle} \qquad (10)$$

Fig.8 gives the SNR value in the experiments to sequentially acquire $\text{TM}^{\text{seq},128\times128}$ and directly evaluated $\text{TM}^{\text{dir},128\times128}$ in which the red lines correspond to the SNR level at the desired output position when each sub-TM were measured. The SNR value level is about 8.9 after $\text{TM}^{\text{sub}}_{Q_1}$ is conducted, then the level increase to 33.95 when $\text{TM}^{\text{sub}}_{Q_2 \text{ with } Q_1 \text{ on}}$ is calculated. As the sequential evaluation procedure goes on, the corresponding SNR keeps climbing due to the increase of the signal level when optimized amplitude distributions are displayed on the corresponding regions on DMD. Eventually, the SNR value reaches 77.12 when $\text{TM}^{\text{sub}}_{Q_1 \text{ with } Q_2, Q_3, Q_4 \text{ on}}$ was conducted. Then it starts to maintain for the rest of the measurement. In contrast, during the direct evaluation of $\text{TM}^{\text{dir},128\times128}$, the SNR level, represented as the blue line, was only 0.9, in which the measured output fields corresponding to input fields would introduce variation due to the low SNR level. It is worth mentioning that even though the captured frames that used to calculate TM is not the same, the total measurement time for both TM evaluation methods doesn`t depend on this parameter since computing vector-matrix multiplication of $\text{TM}^{\text{dir},128\times128}$ costs longer time.

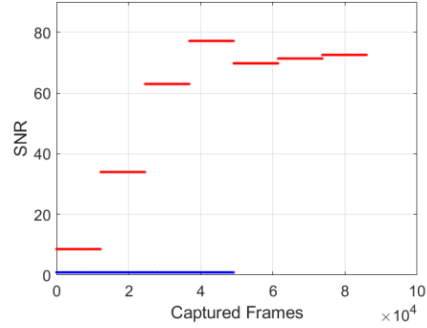

Fig. 8. SNR comparison between direct TM approach and sequential TM approach with size of $128\times128$. Red lines indicate the SNR during sequential measurement as for the sub-TM in size of $64\times64$, each red line then corresponds to $64\times64\times3=12288$ captured signal frames. In contrast, blue line is SNR during the measurement of $TM^{dir,128\times128}$, the corresponding captured frames are $128\times128\times3=49152$.

In our experiments, the processing speed is less affected by the calculation of huge vector-matrix multiplication but mainly limited by the capture framerate of the camera. The Thorlab-CS2100M camera could offer 1200Hz framerate for region of interest $128\times128$ when the exposure time is 0.03 second. Since the DMD can be operated at 22kHz, higher speed operation could be achieved if the framerate of the camera is improved. The PC used in experiment equips with Intel Core (TM) i9-9900K CPU @ 3.60GHz, 64.0GB RAM, 64-bit Operation System, and PM981 Samsung SSD 2048GB. It takes around 10 seconds to record frames and calculate one single $TM^{sub,64\times64}$, the data load to DMD memory costs 15 seconds for each encoded Hadamard basis, so the total measurement time would be approximately 25 minutes for sequential TM evaluation of $TM^{seq,512\times512}$, processing time could be shortened as the image capture speed would be improved by using fast frame camera or photodetector.

## 5. Conclusion

In conclusion, we present a novel sequential TM evaluation approach by carefully characterizing the decomposable nature of the interference effect of transmitted modes through scattering medium. It introduces the advantages of feedback based algorithm that reaches the optimizing distribution while taking into account past evaluation results in a time fashion. The improved SNR brings optimal accuracy of TM evaluation, which leads to a more precise representation of the scattering medium, and further brings superior enhancement factor of focal spot contrast to traditional TM evaluation, in which all linear equations of all parameters are solved together. The proposed approach is experimentally demonstrated for TM size of $128\times128$, $256\times256$ and $512\times512$, which offers researchers an access to fully control elements on modulator without requiring exorbitant computing facility.

**Acknowledgement**

This work was supported by the R&D funding from LinOptx. This work was partially supported by grants from The National Natural Science Foundation of China (61675140).

See Supplement 1 for supporting content

# Supplemental Document

### S1. Description of Input Modes Decomposition

The [Eq. (3)] in main text could be simplified as following formation:

$$I_m^{out} = \left|E_m^{out}\right|^2 = \left(\sum_{i=1}^{N} t_{mi} E_i^{in}\right)^2$$

$$= \left[\sum_{i=1,i\in Q_1}^{N_1} \left(t_{mi} E_i^{in}\right)^2 + \sum_{i=1,i\in Q_1}^{N_1} \sum_{j=1,j\neq i,j\in Q_1}^{N_1} t_{mi} E_i^{in} \cdot t_{mj} E_j^{in}\right.$$

$$\left.+ \sum_{i=1,i\in Q_1}^{N_1} \sum_{j=1,j\in Q_2}^{N_2} t_{mi} E_i^{in} \cdot t_{mj} E_j^{in} + \sum_{i=1,i\in Q_1}^{N_1} \sum_{j=1,j\in Q_3}^{N_3} t_{mi} E_i^{in} \cdot t_{mj} E_j^{in} + \sum_{i=1,i\in Q_1}^{N_1} \sum_{j=1,j\in Q_4}^{N_4} t_{mi} E_i^{in} \cdot t_{mj} E_j^{in}\right] \text{(S1.1)}$$

$$+ \left[\sum_{i=1,i\in Q_2}^{N_2} \left(t_{mi} E_i^{in}\right)^2 + \sum_{i=1,i\in Q_2}^{N_2} \sum_{j=1,j\neq i,j\in Q_2}^{N_2} t_{mi} E_i^{in} \cdot t_{mj} E_j^{in}\right.$$

$$\left.+ \sum_{i=1,i\in Q_2}^{N_2} \sum_{j=1,j\in Q_1}^{N_1} t_{mi} E_i^{in} \cdot t_{mj} E_j^{in} + \sum_{i=1,i\in Q_2}^{N_2} \sum_{j=1,j\in Q_3}^{N_3} t_{mi} E_i^{in} \cdot t_{mj} E_j^{in} + \sum_{i=1,i\in Q_2}^{N_2} \sum_{j=1,j\in Q_4}^{N_4} t_{mi} E_i^{in} \cdot t_{mj} E_j^{in}\right] \text{(S1.2)}$$

$$+ \left[\sum_{i=1,i\in Q_3}^{N_3} \left(t_{mi} E_i^{in}\right)^2 + \sum_{i=1,i\in Q_3}^{N_3} \sum_{j=1,j\neq i,j\in Q_3}^{N_3} t_{mi} E_i^{in} \cdot t_{mj} E_j^{in}\right.$$

$$\left.+ \sum_{i=1,i\in Q_3}^{N_3} \sum_{j=1,j\in Q_1}^{N_1} t_{mi} E_i^{in} \cdot t_{mj} E_j^{in} + \sum_{i=1,i\in Q_3}^{N_3} \sum_{j=1,j\in Q_2}^{N_2} t_{mi} E_i^{in} \cdot t_{mj} E_j^{in} + \sum_{i=1,i\in Q_3}^{N_3} \sum_{j=1,j\in Q_4}^{N_4} t_{mi} E_i^{in} \cdot t_{mj} E_j^{in}\right] \text{(S1.3)}$$

$$+ \left[\sum_{i=1,i\in Q_4}^{N_4} \left(t_{mi} E_i^{in}\right)^2 + \sum_{i=1,i\in Q_4}^{N_4} \sum_{j=1,j\neq i,j\in Q_4}^{N_4} t_{mi} E_i^{in} \cdot t_{mj} E_j^{in}\right.$$

$$\left.+ \sum_{i=1,i\in Q_4}^{N_4} \sum_{j=1,j\in Q_1}^{N_1} t_{mi} E_i^{in} \cdot t_{mj} E_j^{in} + \sum_{i=1,i\in Q_4}^{N_4} \sum_{j=1,j\in Q_2}^{N_2} t_{mi} E_i^{in} \cdot t_{mj} E_j^{in} + \sum_{i=1,i\in Q_4}^{N_4} \sum_{j=1,j\in Q_3}^{N_3} t_{mi} E_i^{in} \cdot t_{mj} E_j^{in}\right] \text{(S1.4)}$$

$$\text{(S1)}$$

### S2. Description of Sequential TM Evaluation Process

The sequential TM evaluation starts from row 1 to row 8 as the procedure operates from left frame (a) to right frame (e) that is shown in Fig.S1. In the most left frame in row 1, blue sqaure indicates the grouped segment region $Q_1$, which projects the encoded Hadamard matrix to front surface of scattering medium like the (b) frame shows, then the sub-TM that corresponds to region $Q_1$ is calculated according to received transmitted light signal and known Hadamard matrix, termed $\text{TM}_{Q_1}^{sub}$, the optimized distribution $\varphi_{Q_1}^{opt}$ is extracted from $\text{TM}_{Q_1}^{sub}$ that is shown in (e) frame, will then be attached on region $Q_1$ as orange square in (a) in row 2, then the Hadamard matrix is now displayed on region $Q_2$ like (b) frame, as a result, the sub-TM $\text{TM}^{\text{sub}}_{Q_2 \text{ with } Q_1 \text{ on}}$ is then measured after calculation, the optimized distribution $\varphi^{\text{opt}}_{Q_2 \text{ with } Q_1 \text{ on}}$ will be induced from $\text{TM}^{\text{sub}}_{Q_2 \text{ with } Q_1 \text{ on}}$. Likewise, the orange square indicates the state that optimized distribution at certain region is displayed, the blue square denotes that certain region is estimated by Hadamard matrix algorithm. The measured optimized distribution from evaluated sub-TM will be displayed during the evaluation of rest regions, as process starts from row 1, the corresponding sub-TMs $\text{TM}^{\text{sub}}_{Q_1}$, $\text{TM}^{\text{sub}}_{Q_2 \text{ with } Q_1 \text{ on}}$, $\text{TM}^{\text{sub}}_{Q_3 \text{ with } Q_1, Q_2 \text{ on}}$, $\text{TM}^{\text{sub}}_{Q_4 \text{ with } Q_1, Q_2, Q_3 \text{ on}}$, $\text{TM}^{\text{sub}}_{Q_1 \text{ with } Q_2, Q_3, Q_4 \text{ on}}$, $\text{TM}^{\text{sub}}_{Q_2 \text{ with } Q_1, Q_3, Q_4 \text{ on}}$, $\text{TM}^{\text{sub}}_{Q_3 \text{ with } Q_1, Q_2, Q_4 \text{ on}}$, $\text{TM}^{\text{sub}}_{Q_4 \text{ with } Q_1, Q_2, Q_3 \text{ on}}$ are measured sequentially until row 8. The eventual optimized distribution that could form focal spot behind scattering medium is constructed by assembling region-based optimized distribution in row 5 to row 8 in terms of their position on input plane, the red cross line shows the border of four sequential evaluated optimized distribution.

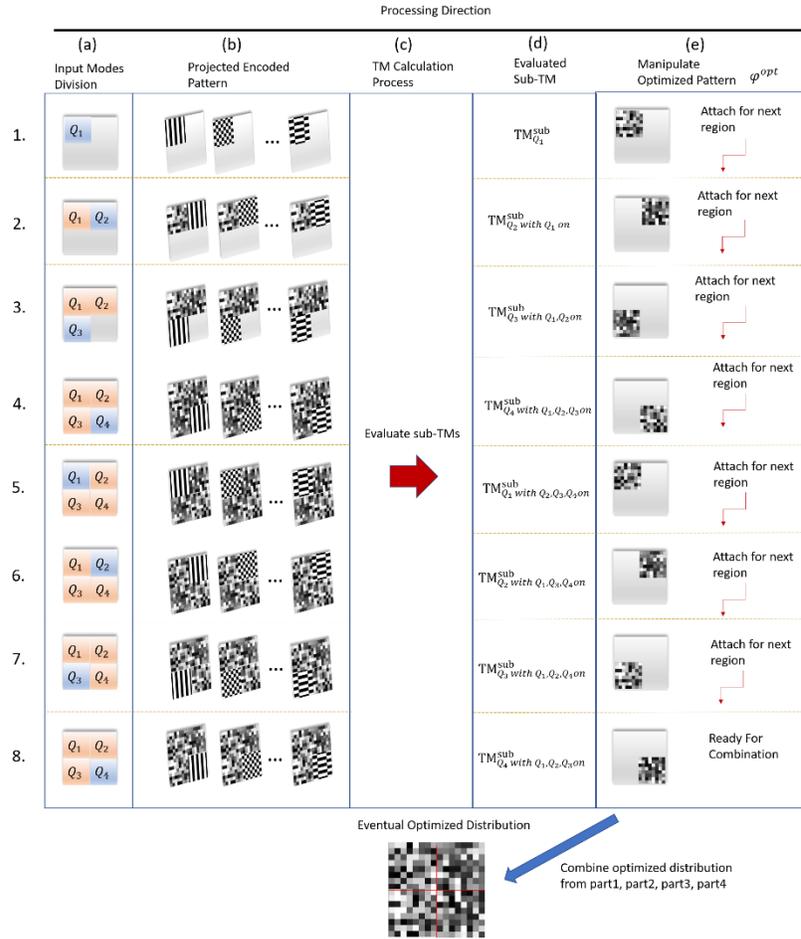

Fig. S1. Process of sequential TM measurement.

## S3. Experiment Result

### 3.1 Maximum Intensity at Desired Output Position

During the sequential TM evaluation of $TM^{seq,128\times128}$, the optimized amplitude distributions that are derived from each sub-TM are displayed on DMD surface regarding to their assigned position region, which would excite point optimization at desired output position. Table S1 shows the maximal intensity value at desired output position, the left column indicates the sub-TM that is measured corresponds to each divided region, and the center column states the symbols of sub-region that are displayed the optimized amplitude distribution. In the right column, it shows the corresponding maximal intensity value when sub-region that is denoted in central column is attached with the optimized distribution, it increases from 1689 to 21171 for the 1$^{st}$ round evaluation, then for the 2$^{nd}$ round evaluation, the intensity continues to increase to 25098 as optimized distribution from regions $Q_1, Q_2$ and $Q_3$ were displayed.

**Table S1 Intensity at output plane correspond to sub-TM**

| Sub-TM Calculation | Optimized Region | Intensity |
|---|---|---|
| $TM^{sub,64\times64}_{Q_1}$ | $Q_1$ | 1689 |
| $TM^{sub,64\times64}_{Q_2 \text{ with } Q_1 \text{ on}}$ | $Q_1, Q_2$ | 8163 |
| $TM^{sub,64\times64}_{Q_3 \text{ with } Q_1, Q_2 \text{ on}}$ | $Q_1, Q_2, Q_3$ | 19025 |
| $TM^{sub,64\times64}_{Q_4 \text{ with } Q_1, Q_2, Q_3 \text{ on}}$ | $Q_2, Q_3, Q_4$ | 21171 |
| $TM^{sub,64\times64}_{Q_1 \text{ with } Q_2, Q_3, Q_4 \text{ on}}$ | $Q_1, Q_3, Q_4$ | 22998 |
| $TM^{sub,64\times64}_{Q_2 \text{ with } Q_1, Q_2, Q_3 \text{ on}}$ | $Q_1, Q_2, Q_4$ | 23541 |
| $TM^{sub,64\times64}_{Q_3 \text{ with } Q_1, Q_2, Q_4 \text{ on}}$ | $Q_1, Q_2, Q_3$ | 25098 |

### 3.2 Sequential TM Evaluation Result in Dimension $256\times256$

The transmission matrix $TM^{seq,256\times256}$ in size of $256\times256$ is also measured by sequential TM evaluation approach in experiment. The size of Hadamard matrix increase to $65536\times65536$, and the corresponding vector-matrix multiplication becomes $\mathbf{S}_{1\times65536}\mathbf{H}_{65536\times65536}$. For sequential TM evaluation method, the surface of DMD region of $256\times256$ is divided to $Q_1, Q_2, ..., Q_{16}$ segment regions. Fig. S2(a) displays the $I_{max}$ at desired output position when the optimized amplitude distributions derived from the previously acquired sub-TMs were attached on DMD, it can be seen that $I_{max}$ is faithfully following the proposed principle and started to increase from 530 after the evaluation of $TM^{sub}_{Q_1}$, to approximately 34840 when 16$^{th}$ sub-TM evaluation was finished, afterwards during the second round evaluation, the intensity value perturbates near 34840 from 17$^{th}$ sub-TM evaluation to 32$^{nd}$ evaluation. Similar to the result for evaluation of $TM^{seq,512\times512}$, the trend of $I_{max}$ experiences three stages of raising rate, it can be also explained by the effect of signal level that same as the evaluation of $TM^{seq,512\times512}$. As a result, $TM^{seq,256\times256}$ would be realized with its produced optimized distribution $\varphi^{opt,256\times256}$ that is shown in Fig. S2(b), and the focal spot will be formed in Fig.S2(c) that shows the light focusing at the output plane behind the scattering medium when $\varphi^{opt,256\times256}$ is attached on the DMD. The logarithm plot of intensity distribution is shown in Fig.S2(d) to visualize the enhancement level.

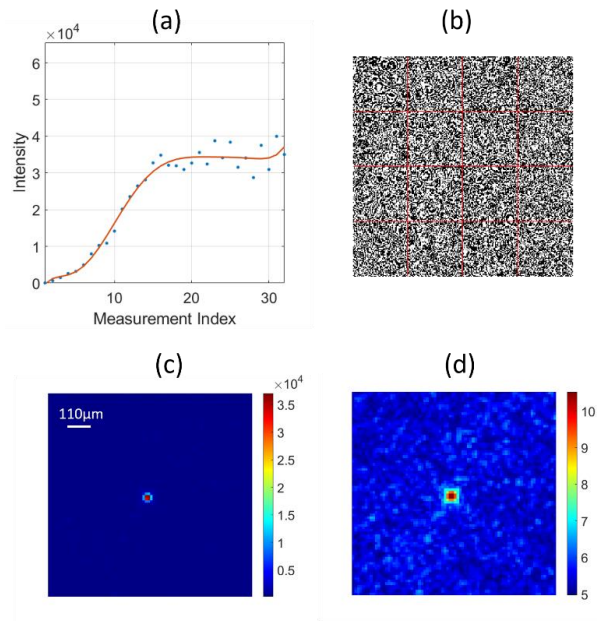

Fig.S2. Experiment result of sequential TM measurement in size of $256 \times 256$. (a) $I_{max}$ at desired position on output plane during sequential TM measurement in 16 regions that generated from $TM^{seq,256\times256}$, blue dots indicates original intensity value, red curve indicates fitting result to original data; (b) Eventual optimized distribution $\varphi^{opt,256\times256}$ from sequential TM approach, the red line indicates border of divided region; (c) Focal spot at desired position when $\varphi^{opt,256\times256}$ is projected to scattering medium; (d) Logarithm distribution of (c).